%% file: Template.tex
\documentclass{article}
\usepackage{spconf,amsmath,graphicx,hyperref,booktabs,url,color,cancel,multirow}
\usepackage{graphicx}
\usepackage{subcaption}

\usepackage{xcolor}

\title{Note-Level Temporal Grounding of Musical Concepts in Large Audio-Language Models}

\name{Kun Fang$^{1,2}$, Ziyu Wang$^{3,4}$, Ichiro Fujinaga$^{1,2}$}
\address{%
$^{1}$ Schulich School of Music, McGill University, Montréal, QC, Canada\\
$^{2}$ CIRMMT (Centre for Interdisciplinary Research in Music Media and Technology), Montréal, Canada\\
$^{3}$ Courant Institute, New York University, New York, USA\\
$^{4}$ Mohamed bin Zayed University of Artificial Intelligence (MBZUAI), Abu Dhabi, UAE
}

\begin{document}
\ninept
\maketitle

\begin{abstract}
Large audio-language models (LALMs) demonstrate growing music-understanding capabilities, but whether their responses are grounded in acoustic evidence remains unclear. Musical language often involves abstract concepts whose acoustic evidence is difficult to define and evaluate precisely. We introduce \textbf{MusicGroundingBench}, a controlled benchmark of algorithmically generated piano audio with exact symbolic alignment, comprising three-note and two-bar settings. We evaluate two complementary capabilities: \emph{grounding}, which localizes the acoustic evidence for a musical query, and \emph{understanding}, which answers questions about the same excerpts. Our experiments show that cross-modal fine-tuning enables models to learn each capability, but adding grounding supervision does not consistently improve understanding across backbones. We further test whether understanding requires listening through audio-ablation controls that remove or replace the input audio, and use attention analysis to examine whether grounding supervision shifts attention toward note boundaries. Meanwhile, the two evaluated LALMs show limited zero-shot grounding even for basic musical concepts, highlighting grounded music understanding as an important open challenge.
 
\end{abstract}


\begin{figure*}[t]
\centering
\includegraphics[width=0.96\textwidth]{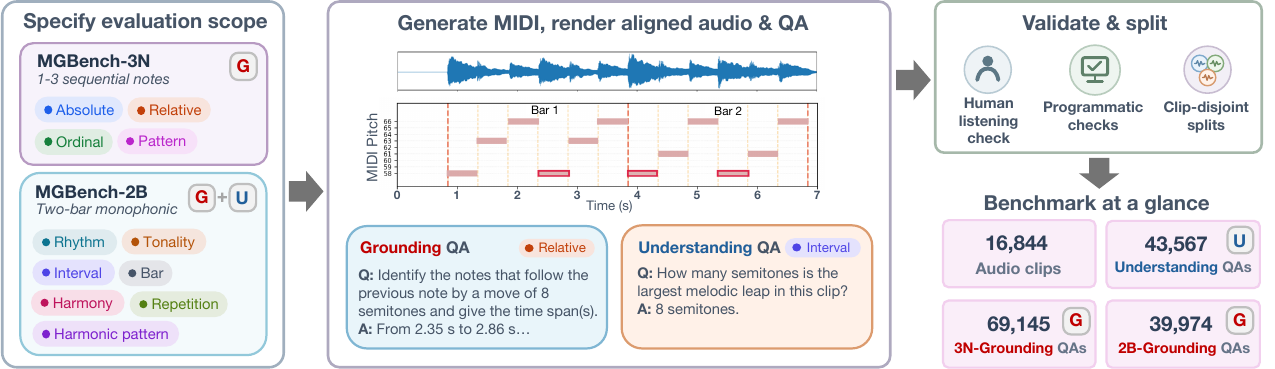}

\caption{
Overview of the MusicGroundingBench construction pipeline.
We first specify the evaluation scope for MGBench-3N and MGBench-2B,
generate symbolic MIDI, render aligned audio, and construct grounding
and understanding QA pairs. The resulting data are then validated and
divided into clip-disjoint splits. G and U denote grounding and
understanding, respectively.
}
\label{fig:benchmark_overview}
\end{figure*}

\section{Introduction}

Large audio-language models (LALMs) are increasingly used for music
captioning, question answering, and instruction following
\cite{gardner_llark_2024, kong_audio_2024, ghosh_music_2025}.
However, evaluations reveal limitations in perceptual understanding
and susceptibility to textual cues
\cite{weck_muchomusic_2024, zang_are_2025}.
Answer accuracy does not tell us how precisely a model can locate
the notes or passages described by a question.

We therefore address localization itself through \textit{note-level temporal music grounding}, where a model receives
an audio clip and a natural-language query and returns the
corresponding time spans. For example,
locating a note reached by a specified interval requires comparing
successive pitches and identifying the destination note's boundaries. Evaluating this ability on real music is difficult, because overlapping
notes, expressive timing, and the lack of note-aligned concept labels
make the acoustic evidence for a musical concept hard to define precisely.

We introduce \textbf{MusicGroundingBench} (MGBench) to support this evaluation through
controlled construction. Algorithmically generated piano MIDI provides explicit musical structure and note timing; reference answers are derived directly from these note events and independently verified.
\textbf{MGBench-3N} evaluates music grounding in clips of up to \textit{three notes}, while
\textbf{MGBench-2B} extends the scope to seven concept groups in \textit{two-bar}
excerpts. The latter provides both grounding and understanding
questions on the same musical material. This synthetic, monophonic
setting provides a simple starting point for testing
whether models can identify the requested notes and locate them
in time.

We first evaluate grounding on both subsets with two zero-shot LALMs,
a text LLM given ground-truth notes, and LLMs
fine-tuned on MGBench. On MGBench-2B, we then fine-tune models for
understanding with and without additional grounding data, and test
whether their answers require listening \cite{zang_are_2025} through
audio-ablation controls. Finally, we use attention analysis to examine
whether grounding supervision shifts attention toward note boundaries
during understanding. Our contributions are:

\begin{itemize}
    \item We introduce \textbf{MusicGroundingBench}, a controlled
    benchmark with note-level references: MGBench-3N for
    grounding and MGBench-2B for grounding and understanding.
    \footnote{The demo, datasets,
    code, and checkpoints are available at \url{https://github.com/kunfang98927/musicgroundingbench}.}

    \item We benchmark grounding with complementary baselines: zero-shot
    LALMs, a text LLM given ground-truth notes, and fine-tuned models.
    \item We propose grounding diagnostics, which test whether
    understanding relies on the audio and whether it benefits from
    grounding supervision.

\end{itemize}

\section{Related Work}
\label{sec:related}

\noindent\textbf{Temporal grounding in audio.}
Localizing natural-language queries to time spans in audio has been
studied as text-to-audio grounding~\cite{xu_text_audio_2021} and
supported by temporally aligned benchmark data~\cite{xie_audiotime_2025}.
Recent work extends this capability to LALMs:
SpotSound~\cite{sun_spotsound_2026} interleaves timestamp tokens with
audio tokens and suppresses hallucinated spans for absent events,
while Audio Flamingo Next~\cite{ghosh_audio_2026} grounds intermediate
reasoning steps to timestamps.
These studies address sound-event localization and temporal reasoning
in general audio; we focus on musical notes and their relations.

\noindent\textbf{Grounding and understanding in music.}
The Audio Flamingo family
\cite{ghosh_audio_2025, ghosh_audio_2025-1, ghosh_music_2025}
advances audio and music understanding, with Music Flamingo
incorporating time-aware audio representations.
MuChoMusic \cite{weck_muchomusic_2024} evaluates music understanding
through multiple-choice QA, while RUListening
\cite{zang_are_2025} reduces reliance on textual cues through
perceptually informed distractors.
SynTheory \cite{wei_music_2024} uses controlled synthetic music
to probe representations of music-theoretic concepts.
MusTBench \cite{kwon_mustbench_2026} evaluates temporal grounding
and reasoning through source boundaries, transitions, event ordering,
and mood trajectories in full-length music.
MusicGroundingBench asks models to locate individual notes and
patterns specified by a query. Reference answers identify the corresponding notes
and their onset and offset times.
We also provide understanding questions on the same clips to
compare music QA performance with and without additional grounding
training.

\section{Benchmark Construction}
\label{sec:benchmark}

MusicGroundingBench comprises two synthetic piano subsets of
increasing complexity; Figure~\ref{fig:benchmark_overview}
summarizes their construction and scale. The suffixes indicate their
musical scope: MGBench-3N contains clips of up to \textit{three notes},
whereas MGBench-2B contains \textit{two-bar} excerpts. The former isolates
note-level localization, while the latter introduces rhythmic, tonal,
intervallic, and harmonic structure and supports both grounding and
understanding questions.
 Across both subsets,
symbolic note events and musical concepts are retained throughout
rendering, enabling exact audio alignment and QA construction. An independent program recomputes answers from the MIDI notes and
checks them against the generated QA pairs.
We test whether answers can be predicted from question text or clip
metadata alone, and revise or remove questions with such
cues.
Additional checks cover answer distributions, formatting, note
boundaries, monophony, and consistency between labels and notes. We also conduct a two-rater listening check on 314 audio--MIDI--QA
items before finalizing the dataset.

\subsection{MusicGroundingBench-3N}
\label{subsec:three_note_benchmark}
\noindent\textbf{Musical material.}
Each example is a 1.5--5.0 s monophonic piano clip of one to three
MIDI-generated notes, rendered with Pianoteq 9 Stage.\footnote{A
physically modeled virtual instrument developed by Modartt.}
The subset contains 6,000 clips spanning all 88 piano keys, with
one-, two-, and three-note clips approximately balanced. Two-note
clips cover unisons through octaves, while three-note clips cover four
contours (\emph{up-up}, \emph{down-down}, \emph{up-down}, and
\emph{down-up}).

\noindent\textbf{QA construction.}
Grounding queries are organized into four families:
ABS (absolute attributes, e.g., ``Locate the note(s) whose pitch is
72 (MIDI)''),
REL (relative comparisons, e.g., ``Locate the pair of notes separated
by 7 semitones''),
ORD (ordinal references, e.g., ``Locate the second note''), and
PAT (pattern-based relations, e.g., ``Locate the notes forming an
ascending step'').
Reference time spans are derived from the annotated onsets and offsets
of the target note events. The subset contains 69,145 grounding QA pairs.

\input{tables/table2-new}

\input{tables/table3-new-unconditional-acc}

\subsection{MusicGroundingBench-2B}
\label{sec:two_bar_benchmark}

\noindent\textbf{Musical material.}
Each sample is a monophonic two-bar piano excerpt generated
algorithmically from symbolic rules and rendered from MIDI. The subset
contains 10,844 clips. Generation controls include meter, tempo,
tonal mode, tonic pitch class, register, pitch-interval tendencies,
bar-level harmonic content, repetition type, and harmonic-pattern type. These controls are realized as note events, for which we retain
onset, offset, pitch, and velocity, together with musical concept labels
derived from the finalized sequence.
\noindent\textbf{QA construction.}
Grounding and understanding QA pairs are derived from the same
annotated excerpts and cover seven concept groups: rhythm, tonality,
interval, bar comparison, harmony, repetition, and harmonic pattern.
Grounding questions request time spans, while understanding questions
request musical descriptions, labels, or counts.
In Figure~\ref{fig:benchmark_overview}, the grounding query asks when notes reached by an
eight-semitone leap occur; the understanding query asks for the
largest interval in the same clip.
Because the material is monophonic, harmony labels refer to
generator-defined chord identities realized as sequential tones.
Other answers are computed from the resulting note sequence.

MGBench-2B contains 39,974 grounding and 43,567 understanding
QA pairs. Across both subsets, MusicGroundingBench comprises
16,844 clips and 152,686 QA pairs.

\section{Evaluation Protocol}
\label{sec:evaluation}

\subsection{Grounding Evaluation}
\label{sec:grounding_eval}

Predictions and references are treated as sets of temporal intervals.
We report the instruction-following rate (IFR), measuring whether
a model produces a valid response.
For models prompted or trained to
produce the required time spans, spans are extracted using the
same regular expressions.
For external LALM baselines, whose outputs are often free-form,
we use Qwen2.5-7B-Instruct~\cite{qwen2.5} as a time span extractor.
All extracted spans are evaluated with the same metrics.

To evaluate localization quality, we compare predicted and reference
spans using temporal intersection-over-union (IoU), with greedy
one-to-one matching at thresholds of 0.5 and 0.7.
A predicted span is matched to a reference span only if their IoU
meets the threshold, and each span can participate in at most one match.
We report span-level precision, recall, and F1, treating matched pairs
as true positives, unmatched predicted spans as false positives,
and unmatched reference spans as false negatives.

We also report MergedIoU, which compares the overall time regions
covered by the predicted and reference spans after overlapping spans
within each set are merged.
It measures the IoU of these regions and is less
sensitive to how a prediction is split into individual segments.
For matched pairs, we report onset/offset mean absolute errors
(MAEs) in milliseconds, measuring how accurately the model places
the start/end boundaries of the target spans.

\subsection{Understanding Evaluation}
\label{sec:understanding_eval}

We use Qwen2.5-7B-Instruct as an LLM judge to accommodate paraphrases
and musically equivalent answers, following prior work on LLM-as-a-judge in MIR~\cite{kun_fang_2024_14877499} and
multimodal understanding~\cite{ge-etal-2025-mllm}.
Given the question, prediction, and reference, the judge
assigns binary labels for instruction following and semantic correctness.
IFR is the proportion of predictions that follow the requested answer
format. Accuracy is the proportion of all test queries answered
correctly.
We also report accuracy within each of the seven concepts defined in
Section~\ref{sec:two_bar_benchmark}.

\begin{figure*}[t]
    \centering
    \begin{subfigure}[b]{0.29\textwidth}
        \centering
        \includegraphics[width=\linewidth]{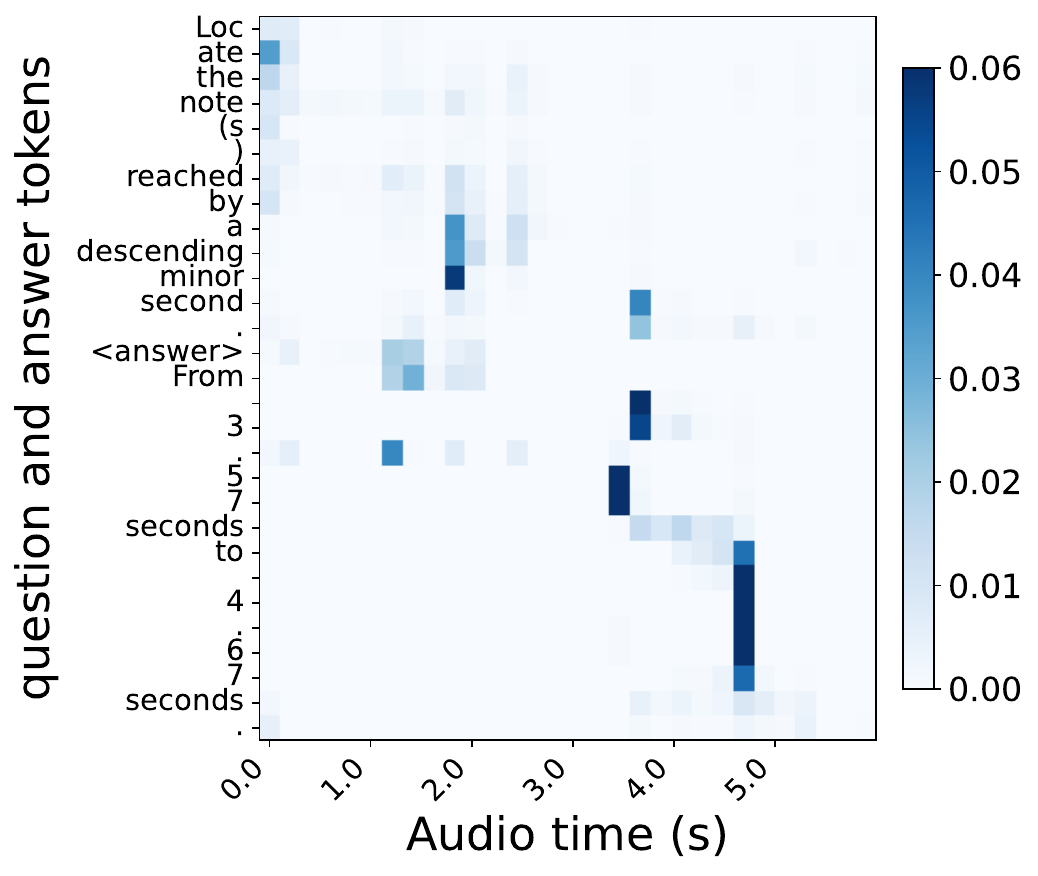}
        \label{fig:attn_heatmap}
    \end{subfigure}
    \hfill
    \begin{subfigure}[b]{0.3\textwidth}
        \centering
        \includegraphics[width=\linewidth]{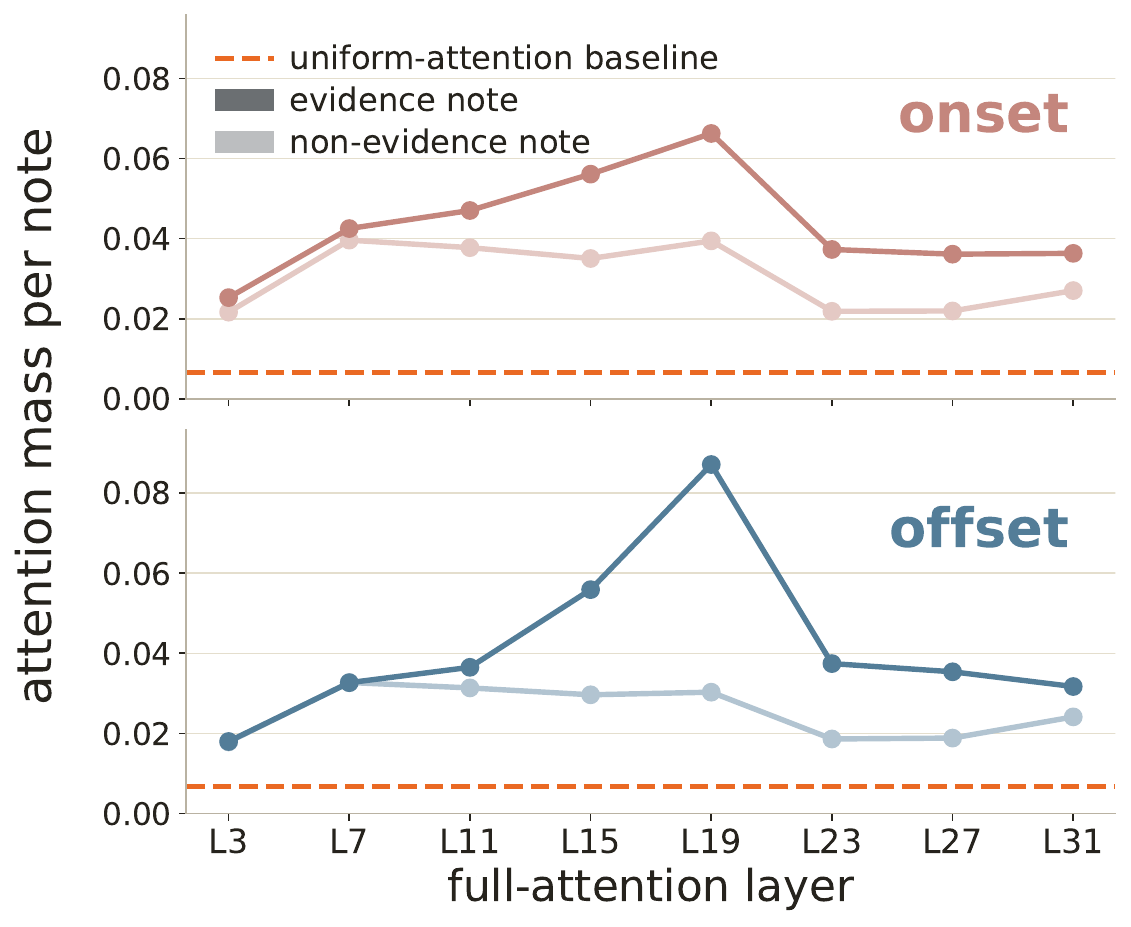}
        \label{fig:grounding_layers}
    \end{subfigure}
    \hfill
    \begin{subfigure}[b]{0.40\textwidth}
        \centering
        \includegraphics[width=\linewidth]{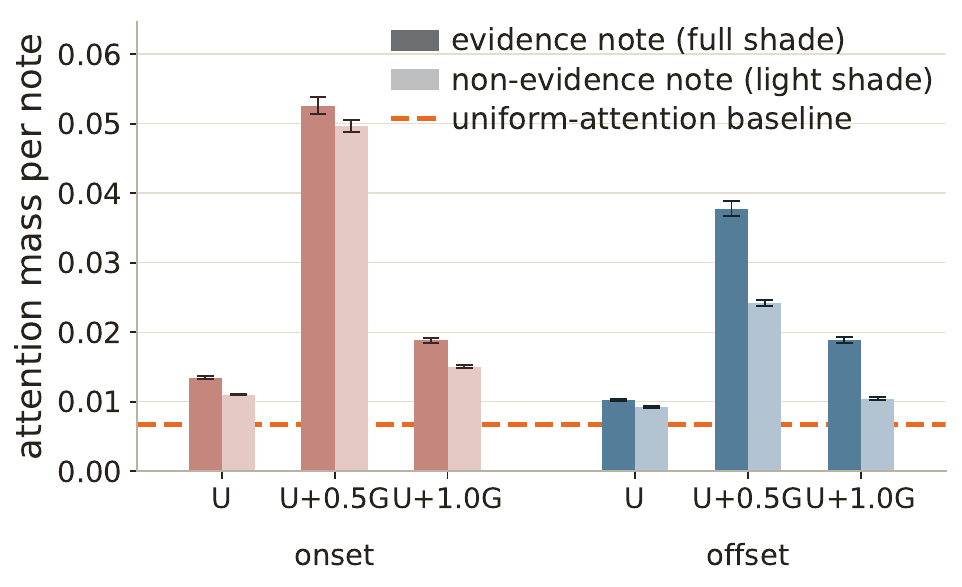}
        \label{fig:understanding_models}
    \end{subfigure}

    \caption{
Attention analysis for Qwen3.5-9B-Base.
Left: An illustrative layer-19 attention map for a grounding query
with the answer 3.57--4.67\,s.
Middle: Attention per note around onsets/offsets across the
grounding model's eight full-attention layers.
Right: The same measure during understanding for $U$, $U+0.5G$,
and $U+1.0G$, averaged over all eight layers.
Full and light shades denote evidence notes needed to determine
the answer and the remaining notes, respectively.
}
    \label{fig:attn_analysis}
\end{figure*}

\section{Experiments}
\label{sec:experiments}

\subsection{Experimental Setup}
\label{sec:exp_setup}

We evaluate four model settings. The FiLM and XAttn proposal baselines~\cite{film,mun_local-global_2020} use MERT features~\cite{li_mert_2024} and trainable text queries and are evaluated only on MGBench-3N. Music Flamingo~\cite{ghosh_music_2025} and Qwen2-Audio-7B-Instruct~\cite{chu2024qwen2audiotechnicalreport} are prompted directly without MGBench training and serve as zero-shot transfer references. The GT-note baseline gives Qwen2.5-7B-Instruct the query and ground-truth MIDI notes without MGBench training; it follows the same timestamp format and evaluation pipeline.

For task-specific models, a two-layer MLP projects frame-level frozen MERT features to the hidden dimension of LLaMA-3-8B~\cite{grattafiori2024llama3herdmodels} or Qwen3.5-9B-Base~\cite{qwen3.5}, and each backbone is adapted with LoRA~\cite{hu_lora_2022}. The models generate time spans for grounding and text for understanding. Both use the same LoRA rank of 16, batch size of 32, learning rate of 2e-4, and five-epoch schedule, with checkpoints selected by validation loss.

For audio-dependence controls, we evaluate the Qwen3.5-9B-Base trained on $U$ with audio removed or replaced by noise, and the unadapted Qwen3.5-9B-Base backbone from question text alone. All conditions use the same questions and scoring protocol.

\subsection{Grounding Results}
\label{sec:main_grounding_results}

Table~\ref{tab:main_results} contrasts zero-shot audio transfer, symbolic-input reasoning, and MGBench-supervised learning. Music Flamingo and Qwen2-Audio transfer poorly, especially at IoU $\geq 0.7$; because neither was trained for our task or output format, they serve as transfer references rather than controlled capability comparisons.

Given exact note identities and boundaries, the GT-note LLM reaches an F1 of 0.5926 on MGBench-3N but only 0.2180 on MGBench-2B at IoU $\geq 0.7$; MergedIoU similarly falls from 0.4353 to 0.1808. Because the notes largely determine the answers, the 3N–2B gap mainly reflects query interpretation and note selection by an untrained LLM; the baseline is not an upper bound.

Despite lower IFR, Qwen3.5-9B-Base localizes more accurately than LLaMA-3-8B. At IoU $\geq 0.7$, their respective F1 scores are 0.8325 versus 0.7629 on MGBench-3N and 0.6752 versus 0.5133 on MGBench-2B. Both models remain substantially weaker on the more structured 2B subset.

\subsection{Understanding Results}
\label{sec:main_understanding_results}

Table~\ref{tab:understanding_results} reports understanding results and input controls. For Qwen trained on $U$, global accuracy drops from 0.6967 with paired audio to 0.2245 without audio and 0.1512 with noise, with declines across all seven concepts. The unadapted text-only backbone reaches 0.2328. Thus, the paired-audio result cannot be explained by question text alone. The zero-shot GT-note Qwen2.5 baseline reaches 0.2718, showing that access to an exact note sequence alone does not make the 2B questions trivial for this text LLM.

Adding grounding produces backbone-dependent outcomes. Qwen improves from 0.6967 under $U$ to 0.7709 and 0.7727 under $U+0.5G$ and $U+1.0G$, whereas LLaMA drops from 0.6113 to 0.4944 and 0.5074. 

For Qwen, gains concentrate in rhythm, interval, and bar comparison; for LLaMA, losses concentrate in tonality and harmony.

\subsection{Attention Analysis}
\label{sec:attn_analysis}

We analyze Qwen3.5-9B-Base attention around note boundaries during
grounding and understanding (Figure~\ref{fig:attn_analysis}).
For answer tokens, we measure the fraction of audio attention within
five-token windows around note onsets or offsets, then divide each
group's fraction by its number of notes.
The middle panel averages 16 heads within each layer; the right
panel averages all eight full-attention layers and heads.
Both then average across samples.
The dashed line gives a uniform-attention reference (the attention expected if it were uniform over time), and error bars
show the standard error of the mean across samples.
The illustrative heatmap (Figure~\ref{fig:attn_analysis} Left) uses all text tokens and the
L19 head with the most attention to reference positions. In the grounding model (Figure~\ref{fig:attn_analysis} Middle), attention to evidence-note boundaries peaks
at L19. During understanding (Figure~\ref{fig:attn_analysis} Right), both mixed-training checkpoints
allocate more attention per note to onset and offset regions than
$U$, for both note groups, even though the questions do not request
timestamps. 
Interestingly, the increase is largest for $U+0.5G$.
The two mixed checkpoints nevertheless have nearly identical QA
accuracy (Table~\ref{tab:understanding_results}), despite their
different levels of boundary attention.

\section{Conclusion}

We introduced MusicGroundingBench for note-level temporal grounding with exact MIDI-derived references and controlled validation. The evaluated LALMs show limited zero-shot transfer, while a zero-shot text LLM given exact notes also struggles on the structured 2B queries. Supervised models learn the task but remain weaker on MGBench-2B than on MGBench-3N. Audio-input controls confirm that the Qwen3.5-based model's understanding depends on paired audio, whereas grounding-augmented understanding remains backbone-dependent and the attention analysis exploratory. Extending the benchmark to polyphonic and real music remains future work.

\vfill\pagebreak

\section{Acknowledgements}

This project is supported in part by the Social Sciences and Humanities Research Council of Canada and the Fonds de recherche du Québec.

\bibliographystyle{IEEEbib}
\bibliography{MusicGrounding}

\end{document}

%% file: tables/table2-new.tex
\begin{table*}[t]
\footnotesize
\centering
\setlength{\tabcolsep}{3.0pt}
\renewcommand{\arraystretch}{0.7}
\begin{tabular}{l cc | cccc | cccc}
\toprule
& & &
\multicolumn{4}{c|}{\textbf{$IoU\geq0.5$}} &
\multicolumn{4}{c}{\textbf{$IoU\geq0.7$}} \\
\cmidrule(lr){4-7}\cmidrule(lr){8-11}
\textbf{Model}
& \textbf{IFR}$\uparrow$
& \textbf{MergedIoU}$\uparrow$
& \textbf{P}$\uparrow$ & \textbf{R}$\uparrow$ & \textbf{F1}$\uparrow$
& \textbf{On./Off. MAE}$\downarrow$
& \textbf{P}$\uparrow$ & \textbf{R}$\uparrow$ & \textbf{F1}$\uparrow$
& \textbf{On./Off. MAE}$\downarrow$ \\
\midrule
\multicolumn{11}{l}{\textbf{MGBench-3N Grounding Test Set} (Count: 900)}\\


GT-Note Baseline  & 0.9433 & 0.4353 & 0.7603 & 0.4979 & 0.6017 & 5.00 / 5.80    & 0.7487 & 0.4903 & 0.5926 & 0.70 / 0.90 \\

Non-LLM (FiLM)
& N/A & 0.4839 & 0.2534 & 0.4176 & 0.3154 & 63.50 / 66.89 & 0.1483 & 0.2444 & 0.1846 & 31.01 / 25.30 \\

Non-LLM (XAttn)
& N/A & 0.5615 & 0.5008 & 0.8296 & 0.6246 & 147.56 / 133.58 & 0.2226 & 0.3687 & 0.2776 & 86.77 / 75.10
\\
Music Flamingo (Zero-shot)

& 0.7867 & 0.0666 & 0.0350 & 0.0243 & 0.0287 & 209.62 / 172.88 & 0.0109 & 0.0076 & 0.0089 & 130.86 / 81.11 \\

Qwen2-Audio-7B-Instruct (Zero-shot)
& 0.0400 & 0.0004 & 0.0000 & 0.0000 & 0.0000 & -- / -- & 0.0000 & 0.0000 & 0.0000 & -- / -- 
\\
LLaMA-3-8B (Ours)
& \textbf{0.9322} & 0.6168 & 0.7741 & 0.8430 & 0.8071 & 22.34 / 35.38 & 0.7317 & 0.7968 & 0.7629 & 18.21 / 30.91 \\
Qwen3.5-9B-Base (Ours)
& 0.9233 & \textbf{0.6784} & \textbf{0.8142} & \textbf{0.8757} & \textbf{0.8439} & \textbf{10.59} / \textbf{21.53} & \textbf{0.8033} & \textbf{0.8640} & \textbf{0.8325} & \textbf{10.06} / \textbf{21.04} \\
\midrule
\multicolumn{11}{l}{\textbf{MGBench-2B Grounding Test Set} (Count: 1035)}\\

GT-Note Baseline & 0.8976 & 0.1808 & 0.2809 & 0.2414 & 0.2597 & 73.50 / 118.40 & 0.2358 & 0.2027 & 0.2180 & 7.90 / 9.00 \\

Music Flamingo (Zero-shot)
& 0.7981 & 0.0389 & 0.0246 & 0.0181 & 0.0208 & 340.80 / 247.49 & 0.0061 & 0.0045 & 0.0052 & 227.68 / 97.47 \\

Qwen2-Audio-7B-Instruct (Zero-shot)
& 0.0077 & 0.0006 & 0.2500 & 0.0010 & 0.0020 & 518.18 / 1174.77 & 0.0000 & 0.0000 & 0.0000 & -- / -- 

\\
LLaMA-3-8B (Ours)
& \textbf{0.9826} & 0.5073 & 0.4666 & 0.6464 & 0.5420 & 20.96 / 27.45 & 0.4419 & 0.6122 & 0.5133 & 13.60 / 19.80 \\

Qwen3.5-9B-Base (Ours)
& 0.9314 & \textbf{0.6001} & \textbf{0.6735} & \textbf{0.7243} & \textbf{0.6980} & \textbf{16.09} / \textbf{19.78} & \textbf{0.6515} & \textbf{0.7007} & \textbf{0.6752} & \textbf{13.14} / \textbf{16.84} \\

\bottomrule
\end{tabular}
\caption{Grounding results on MGBench-3N and MGBench-2B grounding test sets. We report instruction following rate (IFR), MergedIoU computed over the unions of predicted and ground-truth time spans, and precision, recall, F1, onset/offset MAE under greedy matching at IoU thresholds of 0.5 and 0.7, where 0.7 is the stricter criterion. MAE is computed over matched spans only and reported in milliseconds. The GT-note baseline supplies zero-shot Qwen2.5-7B-Instruct with
exact MIDI note identities and boundaries but no audio. External LALMs are evaluated zero-shot without MGBench-specific
training. The remaining models receive task-specific
supervision.}
\label{tab:main_results}
\end{table*}

%% file: tables/table3-new-unconditional-acc.tex
\begin{table*}[t]
\footnotesize
\centering
\setlength{\tabcolsep}{1.8pt}
\newcommand{\nhead}[2]{\textbf{#1}\,{\tiny(#2)}}
\renewcommand{\arraystretch}{0.7}

\begin{tabular}{lc|cccccccc}
\toprule
&
\multicolumn{1}{c|}{} &
\multicolumn{1}{c}{\textbf{Accuracy$\uparrow$}} &
\multicolumn{7}{c}{\textbf{Per-Concept Accuracies$\uparrow$}} \\
\cmidrule(lr){3-3}\cmidrule(lr){4-10}

\textbf{Model / Setting}
&
\textbf{IFR}$\uparrow$
&
\nhead{Global}{1078}
&
\nhead{Rhythm}{161}
&
\nhead{Tonality}{161}
&
\nhead{Interval}{160}
&
\nhead{BarComp.}{160}
&
\nhead{Harmony}{138}
&
\nhead{Repetition}{160}
&
\nhead{HarmonicPat.}{138} \\
\midrule

\multicolumn{10}{l}{\textbf{LALM Baselines (Zero-Shot)}}\\

Music Flamingo
& 0.7208
& 0.2542
& 0.3106
& 0.1553
& 0.1875
& 0.1938
& 0.3116
& 0.2313
& 0.4203 \\

Qwen2-Audio-7B-Instruct
& 0.5779
& 0.1558
& 0.1180
& 0.1739
& 0.1000
& 0.1625
& 0.1232
& 0.1500
& 0.2754 \\

\midrule

\multicolumn{10}{l}{\textbf{LLaMA-3-8B (Ours)}}\\

Train on $U$
& 0.8989
& 0.6113
& 0.4969
& 0.5963
& 0.4313
& 0.4625
& 0.7391
& 0.6875
& 0.9275 \\

Train on $U+0.5G$
& 0.8488
& 0.4944
& 0.5528
& 0.2733
& 0.3938
& 0.4500
& 0.3333
& 0.5688
& 0.9275 \\

Train on $U+1.0G$
& 0.8590
& 0.5074
& 0.5280
& 0.3230
& 0.3500
& 0.5063
& 0.3261
& 0.6125
& 0.9420 \\

\midrule

\multicolumn{10}{l}{\textbf{Qwen3.5-9B-Base (Ours)}}\\

Train on $U$
& 0.9230
& 0.6967
& 0.5714
& \textbf{0.8634}
& 0.4563
& 0.5563
& \textbf{0.7971}
& 0.7313
& 0.9493 \\

Train on $U+0.5G$
& 0.9462
& 0.7709
& \textbf{0.7888}
& 0.8385
& 0.5125
& \textbf{0.7438}
& \textbf{0.7971}
& 0.7625
& \textbf{0.9855} \\

Train on $U+1.0G$
& \textbf{0.9490}
& \textbf{0.7727}
& 0.7640
& 0.8199
& \textbf{0.6250}
& 0.7063
& 0.7609
& \textbf{0.8125}
& 0.9420 \\

\midrule

\multicolumn{10}{l}{
\textbf{Audio-Dependence and Symbolic Controls}
}\\

Qwen $U$: no audio 
& 0.7514
& 0.2245
& 0.2795
& 0.1429
& 0.2250
& 0.2250
& 0.1377
& 0.1688
& 0.4058 \\

Qwen $U$: noise 
& 0.6911
& 0.1512
& 0.1242
& 0.0870
& 0.0625
& 0.1750
& 0.1014
& 0.1125
& 0.4275 \\

Qwen backbone: text only
& 0.7449
& 0.2328
& 0.2298
& 0.1366
& 0.2250
& 0.2125
& 0.0870
& 0.2750
& 0.4783 \\

GT-Note Baseline & 0.7597 & 0.2718 & 0.2298 & 0.2857 & 0.1938 & 0.3125 & 0.1522 & 0.2375 & 0.5072 \\



\bottomrule
\end{tabular}

\caption{Understanding results on the MGBench-2B test set.
Accuracy is computed over all test queries. Semantic
correctness of instruction-following outputs is determined by the
LLM judge. The final four rows report audio-input, text-only, and GT-note
controls defined in the experimental setup. Numbers in parentheses in the column headers denote the number of
test queries.}
\label{tab:understanding_results}
\end{table*}